\documentclass[letterpaper, 10 pt, journal, twoside]{IEEEtran}

\usepackage{amsmath} 
\usepackage{caption}
\usepackage{amssymb}  
\usepackage{cite}
\usepackage{accents}
\usepackage{physics}
\usepackage{bm}      
\usepackage{tabularx}
\usepackage{graphicx}
\usepackage{svg}
\usepackage{placeins} 
\usepackage{float} 
\usepackage{subfigure}
\usepackage{array}
\usepackage{siunitx}
\usepackage{adjustbox}
\usepackage{xcolor}
\usepackage{pgfplots}
\usepackage{adjustbox}
\usepackage{soul}
\usepackage[ruled,vlined,linesnumbered]{algorithm2e}
\usepackage{multirow}
\usepackage{booktabs}

\makeatletter
\let\NAT@parse\undefined
\makeatother
\usepackage{hyperref}

\usepackage{makecell} 

\newcommand{\vecx}{\mathbf{x}}

\newcommand{\vecu}{\mathbf{u}}

\newcommand{\vecxi}{\bm{\xi}}

\newcommand{\matM}{\bm{\tau}}

\newcommand{\inertia}{\mathbf{J}}

\newcommand{\loadmass}{m_{\loadsmall}}
\newcommand{\loadquad}{m_{\quadsmall}}

\newcommand{\angvel}{\bm{\omega}}
\newcommand{\angacc}{\dot{\angvel}}

\newcommand{\robotpos}{\vecx^{\worldfsmall}_{\quadsmall}}
\newcommand{\robotquat}{\mathbf{q}_{\worldfsmall \robotfsmall}}

\newcommand{\robotquatdot}{\dot{\mathbf{q}}_{\worldfsmall \robotfsmall}}
\newcommand{\matR}{\mathbf{R}(\robotquat)}
\newcommand{\matRT}{\mathbf{R}^{\top}(\robotquat)}
\newcommand{\robotrot}{\matR}

\newcommand{\robotvel}{\dot{\vecx}^{\worldfsmall}_{\quadsmall}}
\newcommand{\robotacc}{\ddot{\vecx}^{\worldfsmall}_{\quadsmall}}
\newcommand{\robotangvel}{\bm{\omega}^{\robotfsmall}}
\newcommand{\robotangacc}{\angacc^{\robotfsmall}}

\newcommand{\robotposdes}{\vecx^{\worldfsmall}_{\quadsmall,\refsmall}}
\newcommand{\robotveldes}{\dot{\vecx}^{\worldfsmall}_{\quadsmall, \refsmall}}
\newcommand{\robotquatdesired}{\mathbf{q}_{\worldfsmall \robotfsmall, \refsmall}}
\newcommand{\matRd}{\mathbf{R}(\robotquatdesired)}
\newcommand{\matRdT}{\mathbf{R}^{\top}(\robotquatdesired)}

\newcommand{\robotangveldes}{\bm{\omega}^{\robotfsmall}_{\refsmall}}

\newcommand{\errorposquad}{\mathbf{e}_{\robotpos{}}}
\newcommand{\errorvelquad}{\mathbf{e}_{\robotvel{}}}
\newcommand{\errorattitudequad}{\mathbf{e}_{\mathbf{R}}}
\newcommand{\errorangularquad}{\mathbf{e}_{\robotangvel}}
\newcommand{\robotangvelc}{\bm{\omega}^{\camerafsmall}}

\newcommand{\camerapos}{\vecx^{\robotfsmall}_{\camerasmall}}
\newcommand{\cameraquat}{\mathbf{q}_{\robotfsmall \camerafsmall}}
\newcommand{\cameraquattotal}{\mathbf{q}_{\camerafsmall \worldfsmall}}
\newcommand{\cameraquatinverse}{\mathbf{q}_{\camerafsmall \robotfsmall}}
\newcommand{\matRcw}{\mathbf{R}(\cameraquattotal)}
\newcommand{\rotcw}{\matRcw}
\newcommand{\rotcb}{\mathbf{R}(\cameraquatinverse)}
\newcommand{\rotbc}{\mathbf{R}(\cameraquat)}

\newcommand{\loadpos}{\vecx^{\worldfsmall}_{\loadsmall}}
\newcommand{\loadposdes}{\vecx^{\worldfsmall}_{\loadsmall,\refsmall}}

\newcommand{\loadvel}{\dot{\vecx}^{\worldfsmall}_{\loadsmall}}
\newcommand{\loadveldes}{\dot{\vecx}^{\worldfsmall}_{\loadsmall, \refsmall}}
\newcommand{\loadacc}{\ddot{\vecx}^{\worldfsmall}_{\loadsmall}}

\newcommand{\errorposload}{\mathbf{e}_{\loadpos}}
\newcommand{\errorvelload}{\mathbf{e}_{\loadvel}}

\newcommand{\loadposc}{\vecx^{\camerafsmall}_{\loadsmall}}
\newcommand{\loadposb}{\vecx^{\robotfsmall}_{\loadsmall}}
\newcommand{\loadvelc}{\dot{\vecx}^{\camerafsmall}_{\loadsmall}}
\newcommand{\loadvelb}{\dot{\vecx}^{\robotfsmall}_{\loadsmall}}

\newcommand{\cablevec}[1]{\vecxi^{\worldfsmall}_{#1}}
\newcommand{\cablevecestimate}[1]{\hat{\vecxi}^{\worldfsmall}_{#1}}

\newcommand{\cabledotvec}[1]{\dot{\vecxi}^{\worldfsmall}_{#1}}
\newcommand{\cabledotvecestimate}[1]{\dot{\hat{\vecxi}}^{\worldfsmall}_{#1}}

\newcommand{\cableddotvec}[1]{\ddot{\vecxi}^{\worldfsmall}_{#1}}

\newcommand{\realnum}[1]{\mathbb{R}^{#1}}
\newcommand{\SOthree}{SO(3)}

\newcommand{\twonorm}[1]{\left\lVert#1\right\rVert_2}

\DeclareMathOperator*{\argmin}{min}

\newcommand\scalemath[2]{\scalebox{#1}{\mbox{\ensuremath{\displaystyle #2}}}}
\newcommand{\worldf}{\mathcal{W}}
\newcommand{\worldfsmall}{\scalemath{0.45}{\mathcal{W}}}
\newcommand{\loadsmall}{\scalemath{0.6}{{L}}}
\newcommand{\quadsmall}{\scalemath{0.6}{{Q}}}
\newcommand{\refsmall}{\scalemath{0.6}{{ref}}}

\newcommand{\camerasmall}{\scalemath{0.6}{{C}}}
\newcommand{\robotf}{\mathcal{Q}}
\newcommand{\robotfsmall}{\scalemath{0.45}{\mathcal{Q}}}
\newcommand{\camerafsmall}{\scalemath{0.45}{\mathcal{C}}}

\newcommand{\axis}[2]{\mathbf{e}_{#1}^{#2}}
\newcommand{\cameraf}{\mathcal{C}}
\newcommand{\hamiltomplus}{\bm{\Lambda}^{\scalemath{0.45}{{+}}}}

\newcommand{\prths}[1]{\left(#1\right)}

\newcommand{\shortunderline}[1]{\underaccent{\bar}{#1}}

\title{\LARGE \bf
ES-HPC-MPC: Exponentially Stable Hybrid Perception Constrained MPC for Quadrotor with Suspended Payloads
}

\author{Luis F. Recalde$^{1*}$, Mrunal Sarvaiya$^{2*}$, Giuseppe Loianno$^{2}$, and Guanrui Li$^{1}$
\thanks{Accepted to IEEE Robotics and Automation Letters.}
\thanks{$^*$These authors contributed equally.}
\thanks{$^1$The authors are with the Worcester Polytechnic Institute, Robotics Engineering, Worcester, MA 01609, USA. {\tt\footnotesize email: \{lfrecalde, gli7\}@wpi.edu}.}
\thanks{
$^2$The authors are with the University of California Berkeley,
Department of Electrical Engineering and Computer Sciences,
Berkeley, CA 94720, USA. {\tt\footnotesize email: \{mrunaljsarvaiya, loiannog\}@eecs.berkeley.edu}.}
\thanks{This work was supported by the NSF CPS Grant CNS-2121391, the NSF CAREER Award 2145277, and the DARPA YFA Grant D22AP00156-00}
}

\begin{document}
\makeatletter

\g@addto@macro\@maketitle{
\setcounter{figure}{0}
\centering
    \includegraphics[width=\textwidth]{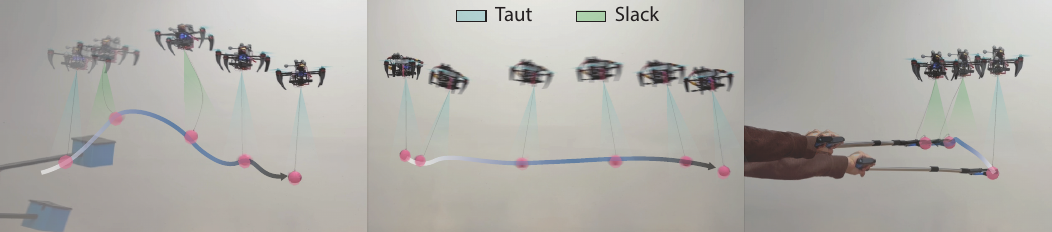} 
    \captionof{figure}{Trajectory tracking experiment considering that the payload is subjected to unexpected hybrid mode
transitions (\textbf{left}). The FoV during hybrid mode transition is represented by the cones in green and blue. Infeasible trajectory that potentially induces the payload
to exit the FoV. The system yaws, increasing the camera’s visibility along the trajectory (\textbf{center}). A human-payload manipulation experiment in which a gripper is used to enforce hybrid mode transitions (\textbf{right}).
\label{fig:cover_image}}
  \vspace{-25pt}
}
\makeatother
\maketitle

\maketitle
\begin{abstract}
Aerial transportation using quadrotors with cable-suspended payloads holds great potential for applications in disaster response, logistics, and infrastructure maintenance. However, their hybrid and underactuated dynamics pose significant control and perception challenges.
Traditional approaches often assume a taut cable condition, limiting their effectiveness in real-world applications where slack-to-taut transitions occur due to disturbances. We introduce ES-HPC-MPC, a model predictive control framework that enforces exponential stability and perception-constrained control under hybrid dynamics. Our method leverages Exponentially Stabilizing Control Lyapunov Functions (ES-CLFs) to enforce stability during the tasks and Control Barrier Functions (CBFs) to maintain the payload within the onboard camera’s field of view (FoV). We validate our method through both simulation and real-world experiments, demonstrating stable trajectory tracking and reliable payload perception. We validate that our method maintains stability and satisfies perception constraints while tracking dynamically infeasible trajectories and when the system is subjected to hybrid mode transitions caused by unexpected disturbances. 
\end{abstract}

 \vspace{-15pt}
\section*{Supplementary material}
Page:
\url{https://lfrecalde1.github.io/es-hpc-mpc.github.io/}
\section{Introduction}
\IEEEPARstart{M}{icro} Aerial Vehicles (MAVs) equipped with onboard sensors are ideal platforms for autonomous navigation and have considerable potential to assist human beings in hazardous operations. They have been applied to search and rescue \cite{Lai_2023}, infrastructure inspection \cite{trujillo2019}, and material transportation \cite{8276269}, with object transport especially valuable for rapid deployment tasks such as disaster response \cite{BARMPOUNAKIS2016ijtst} and installing equipment in inaccessible areas \cite{Lindsey2012ConstructionWQ}.

Aerial transportation systems often employ active or passive manipulation mechanisms \cite{9462539}. Active mechanisms, such as robotic arms and grippers \cite{Suarez2024Through-WindowScenario}, provide precise manipulation capabilities at the cost of increased hardware complexity, inertia, and power consumption. In contrast, passive attachments such as cables are simpler in design and allow for more agility \cite{10243043}. The intrinsic underactuation in cable-driven systems poses challenges for system stability and precise payload tracking, especially during slack-to-taut transitions. Cables are lightweight, passive mechanisms that maintain a safe distance between the aerial robot and the payload, enhancing safety during human interaction. Cables offer a favorable trade-off, effectively balancing load maneuverability, and safety.

In recent years, quadrotors equipped with suspended payloads have become increasingly important in logistics, transportation, and disaster relief \cite{Jimenez-Cano2022PreciseTechniques}. Researchers have developed control and perception methods for this system by leveraging non-linear geometric control \cite{guanrui2021ral, 7892030}, MPC \cite{10156031, 10341785} and computer vision~\cite{sarah2018icra,guanrui2021pcmpc}. However, the aforementioned control and perception strategies are all based on a strong assumption that the cable is always taut. When this assumption breaks due to unexpected disturbances in real-world settings, these strategies fail and potentially lead to system failures \cite{mrunal2025hpampc}.
Distinguishing between the taut and slack phases of a suspended load system is particularly relevant in various scenarios, including: (i) human–robot interaction, where impulsive disturbances from a user can trigger mode transitions; (ii) outdoor environments, where wind or unexpected interactions may induce slack–taut switching; and (iii) transportation of materials, where such transitions may occur during the release or repositioning of the payload. 
HPA-MPC \cite{mrunal2025hpampc} includes a relevant comparison, where they demonstrate that if a controller does not account for hybrid dynamics during a mode transition, the controller's performance degrades, thereby increasing the likelihood of a crash. 
A few recent works \cite{mrunal2025hpampc,haokun2024impact} have explored accounting for the system's hybrid dynamics; however, their controller's stability and convergence properties remain unexplored.

To address these limitations, we propose the ES-HPC-MPC framework for a quadrotor system with a cable-suspended payload, which enforces both trajectory tracking stability and payload visibility across the hybrid cable modes, without compromising control performance. To the best of our knowledge, this is the first work that enforces stability and ensures payload's visibility for quadrotors with suspended payloads while explicitly considering the full hybrid dynamics. We summarize the novel contributions of this work as follows: 
\begin{itemize}
\item We propose dynamically updated Exponentially Stabilizing Control Lyapunov Functions (ES-CLFs) explicitly designed for the hybrid dynamics of quadrotors with cable-suspended payloads, enforcing stable trajectory tracking across both taut and slack cable modes.

\item We leverage Control Barrier Functions (CBFs) to enforce payload visibility within the camera’s field of view in both the taut and slack cable modes, without compromising the trajectory tracking performance.

\item We experimentally validate our proposed method through both simulation and extensive real-world experiments using onboard sensors and computation only. We show that our method outperforms the existing baseline and ensures perception safety and controller stability during trajectory tracking, even when the system experiences unexpected hybrid mode transitions.
\end{itemize}

\section{Related Works} \label{sec:related_works}

Numerous works have been proposed to address the challenges associated with controlling quadrotor cable-suspended load systems by leveraging coordinate-free dynamics based on Lie groups and MPC frameworks \cite{sarah2018icra, 10243043}. However, these methods typically depend on external motion tracking systems and assume that the cable remains taut. Consequently, they are unsuitable for hybrid dynamics involving slack-to-taut transitions and impractical for real-world transportation tasks.

Perception-aware approaches such as PCMPC \cite{guanrui2021pcmpc} enable full-state estimation using on-board sensors; however, they do not account for the system's hybrid dynamics. Furthermore, as discussed in \cite{9483029,9536448}, incorporating the non-linear perception constraints into the MPC increases the likelihood of solver infeasibility.

The approaches in \cite{mrunal2025hpampc, haokun2024impact} are closely related to this work, as they incorporate hybrid dynamics and perception awareness to maintain payload visibility within the camera’s FoV. While \cite{mrunal2025hpampc} introduces a hybrid-aware MPC, it relies on a simple cost function for each hybrid mode and incorporates perception-awareness through a soft cost. This formulation can lead to conservative behaviors since the trajectory tracking and perception objectives are prioritized through their relative cost weights. This trade-off complicates MPC tuning and can adversely affect tracking performance. 
Furthermore, although the framework in \cite{mrunal2025hpampc} addresses both hybrid dynamics and payload visibility, it lacks any convergence properties during trajectory tracking and human–payload interactions. 

Control Lyapunov Functions and CBFs have been successfully applied in robotics, demonstrating their effectiveness in enforcing stability \cite{6709752, 4639455} and safety constraints \cite{8796030, 9483029}. Building on these concepts, we propose ES-HPC-MPC, which integrates ES-CLFs and CBFs to provide stability and perception guarantees for quadrotors with suspended payloads. Our unified approach enforces perception guarantees by keeping the payload within the FoV, enabling accurate full-state estimation while ensuring reliable trajectory tracking. We experimentally verify our method and show that it maintains controller stability and perception visibility during high-acceleration trajectories and unexpected hybrid mode transitions.

\section{Hybrid System Dynamics} \label{sec:system_dynamics}
\begin{figure}[t]
    \centering
    \includegraphics[width= 0.9\columnwidth]{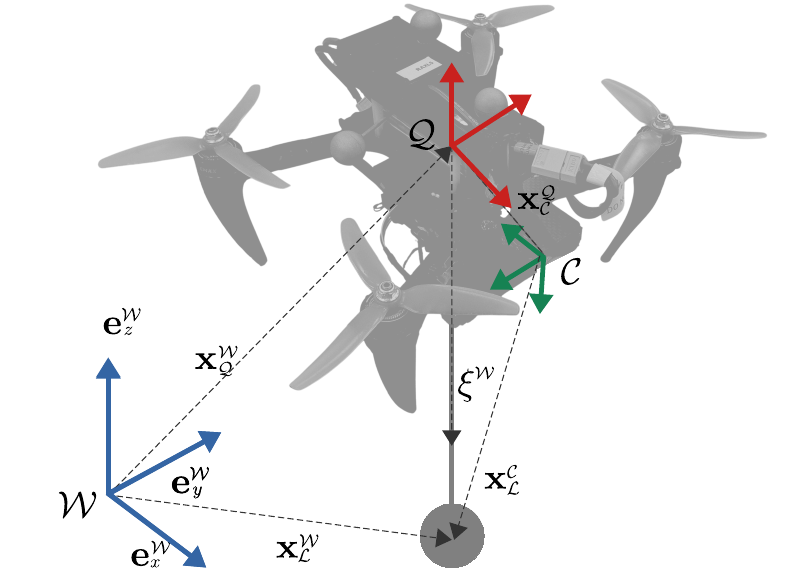} 
    \caption{Quadrotor with a cable-suspended load representation.\label{fig:frame_convention}}
\end{figure}
\begin{table}[t]
    \caption{Notation Table} 
    \label{tab:notation}
    \renewcommand{\arraystretch}{1.2}
    \begin{adjustbox}{width=\columnwidth,center}
    \begin{tabular}{c c} 
       \toprule
         {\bfseries System Variables} & {\bfseries Description} \\ [1ex] 
        \hline 
$\worldf, \robotf, \cameraf$ & World, quadrotor, and camera frames\\
$\robotpos{}, \loadpos{} \in \mathbb{R}^{3}$ & Quadrotor and  payload positions in $\worldf$\\
$\robotvel, \loadvel \in \mathbb{R}^{3}$ & Quadrotor and  payload linear velocities in $\worldf$\\
$\robotquat  \in \mathbb{S}^{3}$ & Unit quaternion from $\robotf$ to $\worldf$  \\ 
$\robotrot{}  \in\SOthree$ & Rotational matrix from $\robotf$ to $\worldf$ \\
$\robotangvel  \in \mathbb{R}^{3}$ & Quadrotor angular velocity in $\robotf$  \\
$ f, f_c \in \mathbb{R}$ &  Collective thrust and cable tension \\
$ \matM \in \mathbb{R}^{3}$ & Torques on the quadrotor in $\robotf$ \\
$\cablevec{}\in S^2$& Unit vector from quadrotor to payload in $\worldf$\\
$\cabledotvec{} \in \mathbb{R}^{3}$& Time derivative of unit vector $\cablevec{}$\\
$\camerapos \in \mathbb{R}^{3}$& Relative position of the camera frame in $\robotf$\\
$\cameraquat  \in \mathbb{S}^{3}$ & Unit quaternion from $\robotf$ to  $\cameraf$ \\
$\loadposc, \loadvelc \in \mathbb{R}^{3}$ & Payload position and 
 linear velocity in $\cameraf$\\
$\robotangvelc  \in \mathbb{R}^{3}$ & Quadrotor angular velocity in $\cameraf$ \\
$ s \in \{0, 1\} $ &  Variable indicating if the cable is taut or slack\\
$\loadquad, \loadmass \in \mathbb{R}_{ > 0}$ & Mass of the quadrotor and the payload\\
$l_{}, g\in\realnum{}$ & Cable length and gravity constant \\
$\mathbf{J} \in \mathbb{R}^{3 \times 3}$ & Inertia matrix of the quadrotor \\
         \bottomrule
    \end{tabular}
    \end{adjustbox}
    \label{table:1} 
    \vspace{-15pt}
\end{table}
\begin{figure*}
\centering
    \includegraphics[width=0.85\textwidth]{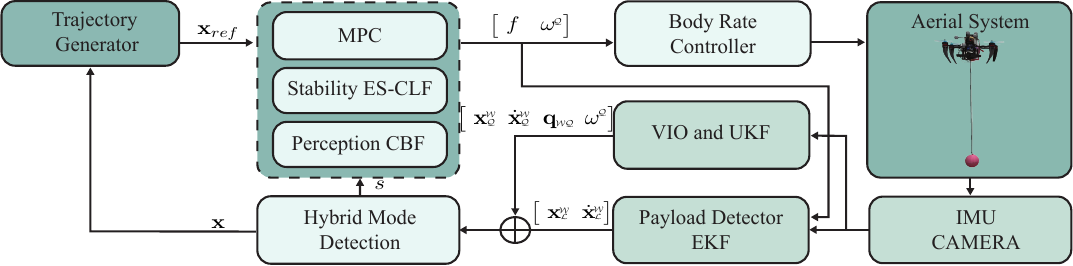}
    \caption{The control block diagram of the proposed ES-HPC-MPC.
    \label{fig:sys_overview}}
    \vspace{-17pt}
\end{figure*}
In this section, we model the hybrid dynamics of a quadrotor transporting a point-mass payload via a massless cable, as shown in Fig.~\ref{fig:frame_convention}. The system exhibits hybrid behavior because the cable can switch between two modes: taut and slack. We introduce a binary switch variable \( s \), where \( s = 1 \) represents the taut mode, and \( s = 0 \) represents the slack mode. We utilize the binary switch variable $s$ to express the quadrotor-payload dynamics in a unified formulation that incorporates this hybrid behavior. This dynamic model forms the foundation for the control and state estimation methods that we develop in Sections~\ref{sec:hpampc} and \ref{sec:state_estimation}.

In the rest of the paper, we represent vectors using boldface notation, where the vector $\mathbf{x}^{\worldfsmall} \in \mathbb{R}^{n}$ is expressed with respect to the world frame $\worldf$. Rotations between coordinate frames are described by unit quaternions, $\robotquat \in \mathbb{S}^{3}$, which defines the orientation of  $\robotf$ respect to $\worldf$.  We denote matrices using boldface uppercase letters, such as \(\mathbf{P} \in \mathbb{R}^{n \times m}\). The \emph{quaternion-to-rotation operator}, $\mathbf{R}(\cdot): \mathbb{S}^{3} \rightarrow SO(3) \subset \mathbb{R}^{3 \times 3}$, maps a unit quaternion to a rotation matrix. The \emph{Hamilton operator}, denoted by $\hamiltomplus(\cdot): \mathbb{S}^{3} \rightarrow \mathbb{R}^{4 \times 4}$, represents quaternion multiplication. Finally, \(\|\cdot\|_{\mathbf{Q}}\) denotes a weighted Euclidean norm, where \(\mathbf{Q} > 0\) is a positive definite matrix. The relevant variables are summarized in Table \ref{tab:notation}. 
\subsection{Quadrotor Dynamics}
As the quadrotor generates thrust and torque to control both its position and orientation, it influences the payload’s motion through the cable when it is taut. 
Using Newton’s laws, we express the quadrotor’s dynamics as follows:
\begin{align}
\robotquatdot &= \hamiltomplus(\robotquat)\robotangvel{},  \frac{d\robotpos{}}{dt} = \robotvel{},\label{eq:hybrid_system_dynamics}\\
\robotacc &=  \frac{f}{\loadquad}\robotrot{}\axis{z}{\worldfsmall} - g \axis{z}{\worldfsmall}  + \frac{f_c}{\loadquad}\cablevec{} s, \label{eq:quad_translation_dynamics}\\
\robotangacc{} &= \inertia^{-1} ( \matM - \robotangvel{}\times\inertia\robotangvel{})\label{eq:quad_rotation_dynamics}, 
\end{align}
where $g= 9.81\,\text{m/s}^2$ is the gravity constant, $\axis{z}{\worldfsmall}$ is the z-axis of the inertia frame.  Note that $f_c$ becomes active only when the cable is taut, i.e. $s=1$. In this taut mode, the quadrotor’s motion is restricted to the surface of a sphere with radius $l$ centered at the payload\cite{guanrui2021pcmpc}. 
We represent this coupling through the unit vector $\cablevec{}$, which points from the quadrotor to the payload, and the cable length $l$. The geometric relationships between the quadrotor and the payload are
\begin{equation} 
\loadpos = \robotpos{} + l\cablevec{}, \quad \loadvel = \robotvel{} + l\cabledotvec{}. \label{eq:quad_load_geometry} 
\end{equation}
To describe the evolution of the cable direction 
$\cablevec{}$, we apply the Lagrange-d’Alembert principle \cite{sreenath2013geometric} to obtain the following equation of cable motion for $\cablevec{}$
\begin{equation}
\cableddotvec{} = \prths{\cablevec{}\times\prths{\cablevec{}\times \robotrot{} \axis{z}{\worldfsmall}\frac{f}{\loadquad l }}} s -\twonorm{\cabledotvec{}}^2\cablevec{} s.\label{eq:cable-dynamics}
\end{equation}
From eqs.~(\ref{eq:quad_translation_dynamics}), (\ref{eq:quad_load_geometry}), and (\ref{eq:cable-dynamics}), we can further derive the tension force in the cable $f_c$ as 
\begin{equation}
f_c = \frac{\loadmass}{\loadmass + \loadquad} \left(l \loadquad (\cabledotvec{}\cdot\cabledotvec{}) - \cablevec{} \cdot (\robotrot{}\axis{z}{\worldfsmall}{f})\right).
\end{equation}

\subsection{Payload Dynamics}
The payload's equations of motion in $\worldf$ can be written as:
\begin{equation}
\frac{d\loadpos}{dt} = \loadvel,~\loadacc = - \frac{f_c}{\loadmass} \cablevec{}s - g \axis{z}{\worldfsmall}.
\end{equation}
As described in Section \ref{subsec:SafePerceptionControlBarrierFunction}, we aim to constrain $\loadposc$ to remain within the camera's FoV, which is computed as follows:
\begin{equation}\label{eq:loadpos_in_cam_fram}
    \loadposc = \rotcw(\loadpos - \robotpos) - \rotcb \camerapos.
\end{equation}
By differentiating  both sides of eq. (\ref{eq:loadpos_in_cam_fram}), we can obtain the payload's velocity with respect to the camera frame:  
\begin{equation}
\loadvelc = \loadposc \times \robotangvelc + \rotcb(\camerapos \times \robotangvel) + \rotcw(\loadvel - \robotvel),
\end{equation}
where $\robotangvelc = \rotcb \robotangvel$.
\subsection{State Space}
The hybrid dynamics of the system can be compactly written in a non-linear control affine form defined as follows:
\begin{equation}
\begin{aligned}
\dot{\vecx} &= \mathbf{f}\prths{\vecx, s} + \mathbf{g}\prths{\vecx, s} \vecu \\
\vecx &= \begin{bmatrix}{\loadpos{}}^{\top}, {\loadvel{}}^{\top}, {\robotpos{}}^{\top}, {\robotvel{}}^{\top}, {\robotquat}^{\top}, {\robotangvel}^{\top}, 
{\loadposc}^{\top} \end{bmatrix}^{\top}
\\
\vecu &= \begin{bmatrix}f,   {\matM}^{\top}\end{bmatrix}^{\top}
\end{aligned}
\label{eq:hybrid_dynamics_control_affine}
\end{equation}
where \(\vecx \in \mathcal{X}\) represents the system's state vector, while $\vecu \in \mathcal{U}$ is the control input.

\section{Exponentially Stable Hybrid Perception Constrained
MPC} \label{sec:hpampc}
In this section, we introduce the \textbf{E}xponentially \textbf{S}table \textbf{H}ybrid \textbf{P}erception-\textbf{C}onstrained \textbf{M}odel \textbf{P}redictive \textbf{C}ontrol (\textbf{ES-HPC-MPC}) for quadrotors transporting suspended payloads. Our method enforces stable trajectory tracking through ES-CLFs and perception safety through CBFs, which enforces the payload's visibility within the camera’s FoV. We highlight that our method has been designed and experimentally validated to work in each hybrid mode. This hybrid-aware formulation enforces controller stability and perception safety, enhancing the system’s resilience to disturbances and improving onboard state estimation reliability. Fig. \ref{fig:sys_overview} provides an overview of the system architecture.

\subsection{MPC Overview}
MPC is a framework designed to compute a sequence of system states and control actions that optimize an objective function while considering the system's dynamics and constraints. Our MPC solves the following optimal control problem (OCP)
\begin{subequations}\label{eq:MPC}
\begin{align}
       \argmin_{\begin{smallmatrix} \vecx, \vecu \end{smallmatrix}} \quad  V(\vecx, \vecx_{ref}, s, T)  + \int_{t_0}^{T} L(\vecx, \vecx_{ref}, s) \, dt,  \label{eq:objective}\\
       \text{subject to } \dot{\vecx} = \mathbf{f}(\vecx, s) + \mathbf{g}(\vecx, s) \vecu,~\forall t \in [t_0, T]  \label{eq:dynamics} \\
       \vecx(t_0) = \vecx_0,   \label{eq:initial_condition} ~~~~~~~~~~~~~~~~~~~~~~\\
       \mathbf{g}_{clf}(\vecx, \vecu, s, \vecx_{ref}) \leq 0,  \label{eq:inequality_constraint_clf}~~~~~~~~~~~~~~~~~\\
       \mathbf{g}_{cbf}(\vecx, \vecu, s) \geq 0, 
       \label{eq:inequality_constraint_cbf} ~~~~~~~~~~~~~~~~~~
\end{align}
\end{subequations}
where \(\vecx_0\) is the initial state and the functions $V$ and $L$ represents the terminal and running  cost respectively.

To ensure exponential convergence of trajectory tracking errors, we design the terminal cost $V$, running cost $L$, and the inequality constraint $\mathbf{g}_{clf}$ based on the ES-CLF described in Section~\ref{sec:Dynamically Update ES-CLF}. Furthermore, to maintain the payload within the camera's FoV, we design the inequality constraint $\mathbf{g}_{cbf}$ based on the CBF formulation shown in Section~\ref{subsec:SafePerceptionControlBarrierFunction}. 

\subsection{Dynamically Updated ES-CLF}\label{sec:Dynamically Update ES-CLF}
In this section, we present the design of the ES-CLF for stable trajectory tracking of the presented hybrid system. To account for the hybrid dynamics, we formulate the Lyapunov function in terms of the mode variable $s$. Specifically, we define the following ES-CLF candidate
\begin{equation}
V(\vecx, \vecx_{ref}, s) =  V_{\robotpos{}} +  V_{\robotquat} +   s V_{\loadpos{}},
\label{eq:lyapunov_translation}
\end{equation}
where $\vecx_{ref}$ is the reference state vector, and $V_{\robotpos{}}$, $V_{\robotquat{}}$, and $V_{\loadpos{}}$ are Lyapunov functions corresponding to the quadrotor’s translational dynamics, rotational dynamics, and the payload’s translational dynamics respectively. We detail the construction of these components in Sections~\ref{subsec:QuadrotorES-CLF} and \ref{Payload ES-CLF}. The binary switch variable $s$, captures the hybrid nature of the system by activating the payload-dependent Lyapunov function only when the cable is taut ($s=1$). This design allows us to accommodate the hybrid dynamics within the stability constraints, ensuring that the ES-CLF remains valid across different modes.

\subsection{Quadrotor ES-CLF}\label{subsec:QuadrotorES-CLF}
\subsubsection{Translational Dynamics}
The translational ES-CLF for the quadrotor can be defined in terms of the position and velocity errors as
\begin{equation}
     \errorposquad = \robotpos{} - \robotposdes,\,\,\,
     \errorvelquad = \robotvel{} - \robotveldes, \label{eq:translation_error_quad}
\end{equation}
where $\errorposquad$ and $\errorvelquad$ denote the quadrotor position and velocity errors, and $\robotposdes$  and $\robotveldes$ represent the desired quadrotor reference position and reference velocity.

Inspired by \cite{10804067}, the Lyapunov function for the quadrotor's translational dynamics is defined as
\begin{equation}
V_{\robotpos{}} = \frac{1}{2} k_q  \| \errorposquad \|^{2} + \frac{1}{2} \loadquad \| \errorvelquad  \|^{2} + c_1 (\errorposquad \cdot \errorvelquad),
\label{eq:lyapunov_translation_quad}
\end{equation}
where $k_q, c_1\in \mathbb{R}^{+}$. 
By applying the Cauchy-Schwartz inequality, $ V_{\robotpos{}}$ can be bounded by
 \begin{equation}
\| \mathbf{z}_{\robotpos{}} \|^{2}_{\shortunderline{\mathbf{Q}}_{1}}  \leq  V_{\robotpos{}} \leq \| \mathbf{z}_{\robotpos{}} \|^{2}_{\bar{\mathbf{Q}}_1},
 \label{eq:lyapunov_translation_quad_inequality}
\end{equation}
where $ \mathbf{z}_{\robotpos{}}  = \begin{bmatrix} \| \errorposquad \|, \| \errorvelquad  \|  \end{bmatrix}^{\top}$ and the matrices $\shortunderline{\mathbf{Q}}_1$ and $\bar{\mathbf{Q}}_1$ are defined as
$$
\shortunderline{\mathbf{Q}}_1 = \frac{1}{2}\begin{bmatrix} k_q & -c_1 \\ -c_1 & \loadquad  \end{bmatrix},\,\,\,\bar{\mathbf{Q}}_1 = \frac{1}{2}\begin{bmatrix} k_q & c_1 \\ c_1 & \loadquad  \end{bmatrix}.
$$
As we show in Section \ref{subsec:mpc_params}, $k_q$ and $c_1$ are selected such that $\shortunderline{\mathbf{Q}}_1$ and $\bar{\mathbf{Q}}_1$ are both positive definite, ensuring that $V_{\robotpos{}} \geq 0$. 

Furthermore, to ensure that $\mathbf{z}_{\robotpos{}}$ is exponentially stable around zero, $\dot{V}_{\robotpos{}}$ must be bounded and satisfy
 \begin{equation}
\dot{V}_{\robotpos{}}(\vecx, \vecx_{ref}, \vecu) \leq - \|  \mathbf{z}_{\robotpos{}} \|^{2}_{{\mathbf{W}}_1},
\label{eq:lyapunov_translation_quad_inequality_dot}
\end{equation}
where ${\mathbf{W}}_1 > 0$ and is tuned as shown in Section \ref{subsec:mpc_params}.

Since eq.~\eqref{eq:lyapunov_translation_quad_inequality} guarantees $V_{\robotpos{}} \geq 0$ and eq.~\eqref{eq:lyapunov_translation_quad_inequality_dot} guarantees $\dot{V}_{\robotpos{}} \leq 0$, $V_{\robotpos{}}$ is a valid ES-CLF ensuring exponential convergence of the tracking errors. 

\subsubsection{Rotational Dynamics}
The rotational ES-CLF for the quadrotor is derived utilizing the attitude $\errorattitudequad$ and angular velocity $\errorangularquad$ errors, which are formulated as
\begin{equation}
     \errorattitudequad = \frac{1}{2}[\matRdT \matR - \matRT \matRd]^{\vee}, 
\end{equation}
\begin{equation}
     \errorangularquad = \robotangvel - \matRT \matRd \robotangveldes\label{eq:attitude_error_quad},
\end{equation}
where $\robotquatdesired$ and $ \robotangveldes$ are the desired attitude and angular velocity respectively.

We define the Lyapunov function for the quadrotor's rotational dynamics as
\begin{equation}
V_{\robotquat} = k_{R} \Psi(\matR, \matRd) + \frac{1}{2} \errorangularquad \cdot  \mathbf{J} \errorangularquad + c_2 (\errorattitudequad \cdot \errorangularquad),
\label{eq:lyapunov_rotational_quad}
\end{equation}
where $k_{R}, c_2 \in \mathbb{R}^{+}$. Since the inertia matrix $\mathbf{J}$ is a positive definite matrix, we can find that $\errorangularquad \cdot \mathbf{J} \errorangularquad$ is bounded by
\begin{equation}
\lambda_m \| \errorangularquad \|^{2} \leq \errorangularquad \cdot \mathbf{J} \errorangularquad \leq \lambda_M \| \errorangularquad \|^{2},
\label{eq:rotational_error_bounds_inertia}
\end{equation}
where $\lambda_m$ and $\lambda_M$ represents the minimum and maximum eigenvalues of $\inertia$ correspondingly. Furthermore, as shown in \cite{LEE2012231}, $\Psi(\mathbf{R}_1, \mathbf{R}_2) = \frac{1}{2} \mathrm{tr} (\mathbf{I} - \mathbf{R}^{\top}_2 \mathbf{R}_1)$ is bounded by
 \begin{equation}
\| \errorattitudequad  \|^{2} \leq \Psi(\matR, \matRd) \leq 2 \| \errorattitudequad  \|^{2}.
 \label{eq:rotational_error_bounds}
\end{equation}

Similar to the analysis in Section~\ref{subsec:QuadrotorES-CLF}, and based on inequalities \eqref{eq:rotational_error_bounds_inertia} and \eqref{eq:rotational_error_bounds}, along with the Cauchy–Schwarz inequality, we find that $V_{\robotquat}$ is bounded by
 \begin{equation}
\| \mathbf{z}_{\robotquat} \|^{2}_{\shortunderline{\mathbf{Q}}_{2}}  \leq  V_{\robotquat} \leq \| \mathbf{z}_{\robotquat} \|^{2}_{\bar{\mathbf{Q}}_2},
 \label{eq:lyapunov_rotational_quad_inequality}
\end{equation}
where $\mathbf{z}_{\robotquat} = \begin{bmatrix} \| \errorattitudequad  \|, \| \errorangularquad \|  \end{bmatrix}^{\top}$  and the matrices $\shortunderline{\mathbf{Q}}_2 > 0$ and $\bar{\mathbf{Q}}_2 > 0$ are
$$
\shortunderline{\mathbf{Q}}_2 = \frac{1}{2}\begin{bmatrix} 2 k_{R} & -c_2 \\ -c_2 & \lambda_m  \end{bmatrix},\,\,\,\bar{\mathbf{Q}}_2 = \frac{1}{2}\begin{bmatrix} 4 k_{R} & c_2 \\ c_2 & \lambda_M  \end{bmatrix}.
$$
Similar to the bounds established for $V_{\robotpos{}}$, we select $k_R$ and $c_2$ such that $\shortunderline{\mathbf{Q}}_2$ and $\bar{\mathbf{Q}}_2$ are positive definite, ensuring that $ V_{\robotquat}$ is positive and a valid Lyapunov function. Furthermore, if $\dot{V}_{\robotquat}$ satisfies
\begin{equation}
\dot{V}_{\robotquat}(\vecx, \vecx_{ref}, \vecu) \leq - \|  \mathbf{z}_{\robotquat} \|^{2}_{{\mathbf{W}}_2},\,\,\,
{\mathbf{W}}_2 > 0 \label{eq:lyapunov_rotational_quad_inequality_dot}
\end{equation}
then $V_{\robotquat{}}$ is a valid ES-CLF, guaranteeing almost global convergence. 

\subsection{Payload ES-CLF}\label{Payload ES-CLF}
Similar to quadrotor ES-CLF, the payload translational ES-CLF is defined in terms of the payload's position errors $\errorposload$ and velocity errors $\errorvelload$
\begin{equation}
     \errorposload= \loadpos{} - \loadposdes,\,\,\,
     \errorvelload= \loadvel{} - \loadveldes,\label{eq:translation_error_load}
\end{equation}
where $\loadposdes$ and $\loadveldes$ are the desired payload position and velocity.
A Lyapunov function for the payload dynamics can be expressed as
 \begin{equation}
 V_{\loadpos{}} = \frac{1}{2} k_l  \| \errorposload \|^{2} + \frac{1}{2} \loadmass \| \errorvelload  \|^{2} + c_3 (\errorposload \cdot \errorvelload),
\label{eq:lyapunov_translation_load}
\end{equation}
where  $k_l, c_3 \in \mathbb{R}^{+}$. Similar to Section \ref{subsec:QuadrotorES-CLF}, by applying Cauchy-Schwartz inequality, we can obtain that
 \begin{equation}
\| \mathbf{z}_{\loadpos{}} \|^{2}_{\shortunderline{\mathbf{L}}_{1}}  \leq  V_{\loadpos{}} \leq \| \mathbf{z}_{\loadpos{}} \|^{2}_{\bar{\mathbf{L}}_1},
 \label{eq:lyapunov_translation_load_inequality}
\end{equation}
where $ \mathbf{z}_{\loadpos{}}  = \begin{bmatrix} \| \errorposload \|, \| \errorvelload  \|  \end{bmatrix}^{\top}$ and the matrices $\shortunderline{\mathbf{L}}_1 > 0$ and $\bar{\mathbf{L}}_1 > 0$ are expressed as
$$
\shortunderline{\mathbf{L}}_1 = \frac{1}{2}\begin{bmatrix} k_l & -c_3 \\ -c_3 & \loadmass   \end{bmatrix},\,\,\,\bar{\mathbf{L}}_1 = \frac{1}{2}\begin{bmatrix} k_l & c_3 \\ c_3 & \loadmass  \end{bmatrix}.
$$

Finally, if $\dot{V}_{\loadpos{}}$ also satisfies
\begin{equation}
\dot{V}_{\loadpos{}}(\vecx, \vecx_{ref}, \vecu) \leq - \|  \mathbf{z}_{\loadpos{}} \|^{2}_{{\mathbf{W}}_3},\,\,\,
{\mathbf{W}}_3 > 0,
\label{eq:payload_lyapunov_derivative}
\end{equation}
then $V_{\loadpos{}}$ is a valid ES-CLF, guaranteeing exponential convergence of the payload tracking errors.

\subsection{Lyapunov Stability and ES-CLF Constraints}
To ensure stability in our hybrid system, we establish bounds on the control Lyapunov function of the whole system $V$, by combining the bounds on the quadrotor and payload control Lyapunov functions. Hence, 
 \begin{equation}
\| \mathbf{z} \|^{2}_{\shortunderline{\mathbf{Q}}} \leq V(\vecx, \vecx_{ref}) \leq \| \mathbf{z} \|^{2}_{\bar{\mathbf{Q}}},
 \label{eq:lyapunov_inequality}
\end{equation}
where $\mathbf{z} = \begin{bmatrix} \mathbf{z}_{\robotpos{}},  \mathbf{z}_{\robotquat}, \mathbf{z}_{\loadpos{}} \end{bmatrix}^{\top}$. The matrices $\shortunderline{\mathbf{Q}}$ and $\bar{\mathbf{Q}}$ are block diagonal positive definite matrices that can be expressed as: $
\shortunderline{\mathbf{Q}} = \mathrm{blkdiag} \{\shortunderline{\mathbf{Q}}_1, \shortunderline{\mathbf{Q}}_2, s \shortunderline{\mathbf{L}}_1 \}$
and $\bar{\mathbf{Q}} = \mathrm{blkdiag} \{  \bar{\mathbf{Q}}_1, \bar{\mathbf{Q}}_2, s \bar{\mathbf{L}}_1 \}
$.

To enforce stability, the time derivative of $V$ must satisfy
\begin{equation}
\dot{V}(\vecx, \vecx_{ref}, \vecu) \leq - \|  \mathbf{z} \|^{2}_{{\mathbf{W}}},
 \label{eq:lyapunov_inequality_dot}
\end{equation}
where $\mathbf{W} = \mathrm{blkdiag} \{ \mathbf{W}_1, {\mathbf{W}}_2, s {\mathbf{W}}_3 \}$.

By satisfying eqs.~\eqref{eq:lyapunov_inequality} and \eqref{eq:lyapunov_inequality_dot}, we enforce $V$ to be a valid ES-CLF for the hybrid quadrotor with suspended payload system. To integrate this into our MPC framework, we first incorporate the ES-CLF into the optimal control problem by defining both the running and terminal cost based on the Lyapunov function. Secondly, we incorporate the inequality eq.~\eqref{eq:lyapunov_inequality_dot} as constraints in our MPC.
This formulation ensures that our MPC controller actively enforces stability throughout the trajectory execution, even during hybrid mode transitions.

\subsection{Perception CBF Constraints}\label{subsec:SafePerceptionControlBarrierFunction}
To maintain payload visibility, we define a CBF that ensures the load remains inside the camera’s FoV throughout the operation. We formulate this perception CBF as
\begin{equation}
{h(\vecx)} = -\|\loadposc\|^{2}_{\mathbf{K}} + r^{2}_{max} \geq 0,
\label{eq:barrier_function_perception}
\end{equation}
where $\mathbf{K} > 0$ characterizes the shape of the ellipse, accounting for the camera’s asymmetric resolution and $r_{max}$ represents the maximum radius of the safety set within the camera frame. 

To satisfy CBF conditions \cite{8796030, 9483029}, we require that the time derivative of 
$h(\vecx)$ satisfies the following constraint
\begin{equation}
\dot{h}(\vecx, \vecu)  \geq - \alpha h(\vecx),
\label{eq:barrier_set_condition_perception}
\end{equation}
where $\dot{h}(\vecx, \vecu)$ is derived using Lie derivatives, and $\alpha > 0$ is a scalar. We incorporate this constraint into our MPC framework to enforce perception safety. 
By ensuring the payload always remains visible, we enhance state estimation robustness and improve the quadrotor’s ability to track dynamically complex trajectories without perception failures.

\section{State Estimation} \label{sec:state_estimation}

We determine the state of the quadrotor in the $\worldf$ frame through VIO and subsequently estimate $\loadpos$ and $\loadvel$ by estimating the payload position in the camera frame.

\subsubsection{Visual Inertial Odometry}
 An Unscented Kalman Filter (UKF) is employed to integrate  Visual-Inertial Odometry (VIO) measurements with accelerometer and gyroscope data obtained from the Inertial Measurement Unit (IMU), providing state prediction updates. For further information on this method, we refer the reader to \cite{guanrui2021pcmpc, mrunal2025hpampc, justin2017ral}.

\subsubsection{Payload State Estimation} \label{subsec:state_estimation_payload}
To estimate the position and velocity of the payload, we extract the payload measurements from the camera $\loadposc$ and project them into the quadrotor frame as $ \loadposb = \camerapos + \rotbc \loadposc $. 

We utilize an extended Kalman filter (EKF) to achieve higher frequency updates by estimating the cable's direction and velocity through the integration of camera measurements, IMU, and the motors thrust, as detailed below
\begin{equation}
    \hat{\vecx} =  \begin{bmatrix} \cablevecestimate{} & \cabledotvecestimate{} \end{bmatrix}^{\top},~ 
    \vecu = \begin{bmatrix}
         f & \robotquat & \robotangvel{}
    \end{bmatrix}^{\top}.
\end{equation}
The process model is defined as
\begin{equation}
\dot{\hat{\vecx}} = \begin{bmatrix}
    \cabledotvecestimate{} \\
    \frac{1}{\loadquad l } \cablevecestimate{} \times (\cablevecestimate{} \times \robotrot{} \axis{z}{\worldfsmall}f) - \left\lVert \cabledotvecestimate{} \right\rVert_2^2 \cablevecestimate{}
\end{bmatrix} + \mathbf{n},
\end{equation} where $f$ is computed by the controller, \( \robotquat \) and \( \robotangvel{} \) are obtained from the VIO and the IMU, respectively, \(\mathbf{n} \in \mathbb{R}^6\) is the process model noises modeled as zero-mean white noise.
The measurement model is defined as 
\begin{equation}
\mathbf{y} = \begin{bmatrix}
    \loadposb \\
    \loadvelb
\end{bmatrix} = \begin{bmatrix}
    \matRT l \cablevecestimate{} \\
    l \left( \matRT \cabledotvecestimate{}  - \robotangvel{} \times \matRT \cablevecestimate{} \right)
\end{bmatrix} + \mathbf{v},
\end{equation}
where \(\mathbf{v} \in \mathbb{R}^6\) is the measurement model noise modeled as zero-mean additive white Gaussian noise. The variable \( \loadvelb \) can be determined using numerical methods.

\subsubsection{Hybrid Mode Detector} \label{sec:mode_detector}
We identify the hybrid modes of the system using the following condition:
\begin{equation}
s =
\begin{cases} 
0 & \text{if } \left\| \loadpos - \robotpos{} \right\| < l - \epsilon\\ \label{eq:slack_taut_condition_2}
1 & \text{else } 
\end{cases}
\end{equation}
where $\epsilon=0.05~\si{m}$ and is experimentally determined. We also use a low-pass filter to attenuate any high frequency noise in our estimation pipeline.

\section{Experimental Results}  \label{sec:results}
We present the results of our proposed method in both simulation and real-world settings using onboard sensors and compute. We demonstrate that our method can ensure stability and safe perception under high acceleration scenarios while tracking trajectories and during unexpected hybrid mode transitions caused by external disturbances.

\begin{figure}[t]
    \includegraphics[width=\columnwidth]{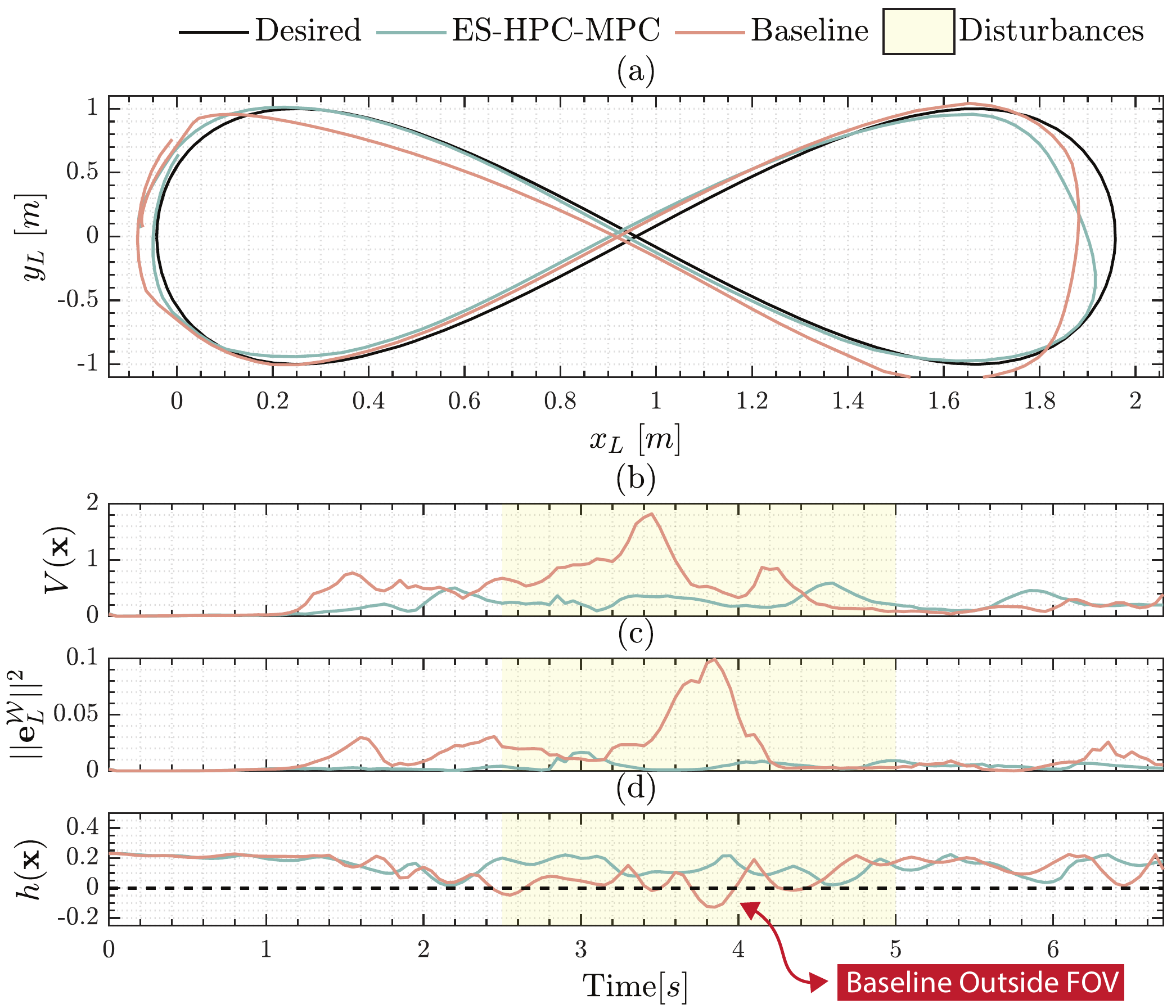}
    \caption{Simulation results for a Lissajous trajectory with speeds up to \(2.4\,\mathrm{m/s}\) under an external step force of \(0.6\,\mathrm{N}\) applied for \(0.5\,\mathrm{s}\). 
 \textbf{a}: Top  view of the simulation using ES-HPC-MPC (blue line) and baseline (orange line).  
 \textbf{b}: ES-CLF values.
 \textbf{c}: Payload position errors.  
 \textbf{d}: CBF values.
    \label{fig:simulation_experiment}}
     \vspace{-15pt}
\end{figure}
Our software is implemented in ROS2 using C++. We run ES-HPC-MPC onboard at $150$ Hz and solve the OCP using acados's \cite{verschueren2022acados} real-time iteration scheme. Furthermore, we incorporate slack variables into the optimization problem to mitigate conflicts between stabilization and safety guarantees.
\vspace{-5pt}
\subsection{MPC Parameters} \label{subsec:mpc_params}
In this section, we detail the tuning parameters for our proposed method. To ensure the exponential convergence of our proposed CLF within our MPC framework, we enforce the derivative of the CLF to satisfy the inequality constraint given in \eqref{eq:lyapunov_inequality_dot}. A key challenge is selecting an appropriate weight matrix \(\mathbf{W}\) for the CLF constraint. 

To address this, we draw inspiration from the geometric control theory \cite{10804067}, which carefully designs the $\mathbf{W}$ matrix to bound the derivatives of the Lyapunov function. We structure the $\mathbf{W}$ matrix in a similar way and apply them in our MPC method as
\begin{equation*}
\begin{split}
\mathbf{W}_1 =\begin{bmatrix} \frac{c_1 k_q}{\loadquad} & -\frac{c_1 k_{vq}}{2 \loadquad} \\ -\frac{c_1 k_{vq}}{2 \loadquad} & k_{vq} - c_1  \end{bmatrix},~{\mathbf{W}}_2 = \begin{bmatrix} \frac{c_2 k_{R}}{\lambda_m} & \frac{c_2 k_{\omega}}{2 \lambda_m} \\ \frac{c_2 k_{\omega}}{2 \lambda_m} & k_{\omega} - \frac{1}{2} c_2  \end{bmatrix},  \\
{\mathbf{W}}_3 =\begin{bmatrix} \frac{c_3 k_l}{\loadmass} & -\frac{c_3 k_{vl}}{2 \loadmass} \\ -\frac{c_3 k_{vl}}{2 \loadmass} & k_{vl} - c_3  \end{bmatrix},\,\,\, \label{eq:lyapunov_translation_load_inequality_dot}
\mathbf{W} = \mathrm{blkdiag} \{ \mathbf{W}_1, {\mathbf{W}}_2, s {\mathbf{W}}_3 \},
\end{split}
\end{equation*}
where $k_{vq}, k_{\omega}, k_{vl}$ are positive constants. We refer the readers to \cite{10804067} for more details of the derivation of the $\mathbf{W}$ matrix and the appropriate selections of $(k_{vq}, k_q, c_1)$, $(k_{R}, c_2, k_{\omega})$ and $(k_l, c_3, k_{vl})$. The parameter $\alpha$ is a positive constant, tuned experimentally.
\subsection{Simulation Experiments}
\begin{table}[t]
\centering
\caption{Simulation Parameters}
\begin{adjustbox}{width=\columnwidth,center}
\renewcommand{\arraystretch}{1.2}
\begin{tabular}{l c l c} 
 \toprule
\toprule
\textbf{System Parameter} & \textbf{Value} & \textbf{Simulation Parameter} & \textbf{Value} \\
\midrule
Quadrotor mass $\loadquad$   & $0.715~\si{kg}$ & Time step $t_s$        & $0.01~\si{s}$ \\
 Payload mass $\loadmass$     & $0.103~\si{kg}$ & Control frequency           & $150~\si{Hz}$ \\
 Cable length $l$             & $0.481~\si{m}$  & Simulation frequency          & $500~\si{Hz}$  \\
Inertia Matrix $\mathbf{J}$             & $0.002 \cdot \mathbf{I}_3$  & Integration method          &  $\text{4\textsuperscript{th}order Runge Kutta}$  \\
\bottomrule
\end{tabular}
\end{adjustbox}
\label{tab:simulation_parameters}
\end{table}

We simulate 20 Lissajous trajectories (1.5–3.5 m/s) to compare our method, ES-HPC-MPC, with the baseline HPA-MPC \cite{mrunal2025hpampc}. We report the root mean squared errors (RMSE) of the trajectory tracking and perception constraint violation rates in Table \ref{table_simulation}, while also evaluating performance under model mismatch and external step force disturbances (ranging from 0.5 to 0.7 $\si{N}$ with a duration of 0.5 $\si{s}$) that trigger hybrid mode transitions. The details of the simulation parameters are included in Table \ref{tab:simulation_parameters}.

\begin{table}[t]
    \centering
    \caption{Comparison between ES-HPC-MPC and the baseline under unmodeled dynamics and external disturbances.}
    \renewcommand{\arraystretch}{1.2}
    \begin{adjustbox}{width=\columnwidth,center}
    \begin{tabular}{lcccccc}
        \toprule\toprule
        & \multicolumn{2}{c}{\bfseries Payload Position RMSE $[m]$}
        & \bfseries Error Reduction
        & \multicolumn{2}{c}{\begin{tabular}{c}\bfseries Perception Constraint\\\bfseries Violations Rate\end{tabular}} \\
        & \multicolumn{2}{c}{mean $\pm$ STD} & \bfseries (\%) & \multicolumn{2}{c}{$h(\vecx)<0$} \\
        \cmidrule(lr){2-3}\cmidrule(lr){4-4}\cmidrule(lr){5-6}
        & ES-HPC-MPC & Baseline & & ES-HPC-MPC & Baseline \\
        \midrule
        \begin{tabular}[l]{@{}l@{}}Ideal\\model\end{tabular}
        & \textcolor{blue}{$0.049 \pm 0.001$} & $0.076 \pm 0.004$
        & $35.5\%$ & $0\%$ & $0\%$ \\
        \midrule
        \begin{tabular}[l]{@{}l@{}}Payload Mass\\Increased by $50\%$\end{tabular}
        & $0.091 \pm 0.003$ & \textcolor{blue}{$0.085 \pm 0.003$}
        & $-7.1\%$ & $0\%$ & $0\%$ \\
        \midrule
        \begin{tabular}[l]{@{}l@{}}External Force\\$0.6\,[\mathrm{N}]$\end{tabular}
        & \textcolor{blue}{$0.052 \pm 0.001$} & $0.120 \pm 0.040$
        & $56.7\%$ & $0\%$ & \textcolor{red}{$10\%$} \\
        \midrule
        \begin{tabular}[l]{@{}l@{}}External Moment\\$0.04\,[\mathrm{N\,m}]$\end{tabular}
        & \textcolor{blue}{$0.049 \pm 0.001$} & $0.0855 \pm 0.013$
        & $42.7\%$ & $0\%$ & \textcolor{red}{$10\%$} \\
        \bottomrule
    \end{tabular}\label{table_simulation}
    \end{adjustbox}
    \vspace{-15pt}
\end{table}

From Table \ref{table_simulation}, we can observe that ES-HPC-MPC achieves better performance in payload tracking errors in almost all experiments, achieving improvements of up to $56.7\%$ relative to the baseline. It also ensures that the payload remains within the FoV under parameter mismatches and external disturbances. In experiments involving an increase in payload mass, the baseline demonstrated a lower RMSE. We attribute this result to the cost function used in \cite{mrunal2025hpampc}, which considers only the tracking error of the payload.  Additionally, the baseline formulation cannot ensure safe perception when disturbances are introduced, causing payload to leave the FoV with a rate of $10\%$ in the 20 trials.

Fig. \ref{fig:simulation_experiment} presents the tracking results for one of the simulated trajectories, including the ES-CLF values, control error metrics, and the CBF values. Our proposed approach outperforms the baseline, particularly in maintaining the payload within the camera’s FoV under external disturbances. When the disturbance is applied (yellow region in Fig. \ref{fig:simulation_experiment}), the baseline method cannot ensure that the payload remains within the camera's FoV, as \(h(\vecx)\) turns negative. 
In contrast, our approach guarantees the payload visibility and exponential stability, even during slack-to-taut transitions.
\subsection{Real World Experiments}
In this section, we experimentally validate our method's capability to enforce stable trajectory tracking while keeping the payload in the camera's FoV. First, we subject the payload to unexpected impulse disturbances that cause hybrid mode switches while tracking straight-line trajectories. Second, we evaluate our method in tracking high acceleration trajectories. And finally, we showcase the advantages of our method in human-payload manipulation scenarios where the system remains slack for long periods of time. 
\begin{figure}[t]
    \includegraphics[width=\columnwidth]{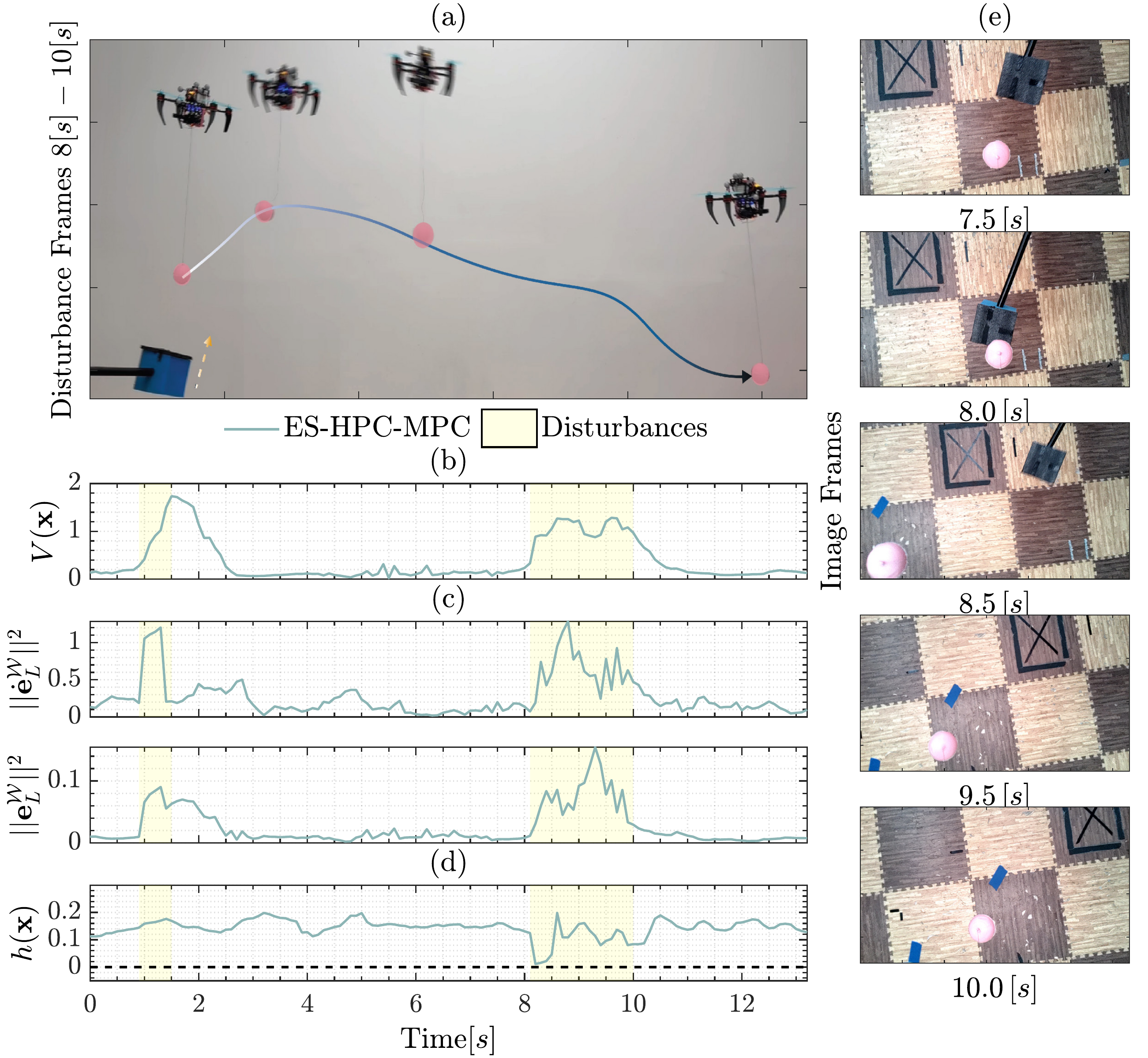}
    \caption{Tracking results for a straight line trajectory with a maximum speed of up to \(2.0\,\mathrm{m/s}\) under vertical external disturbances that generates slack-to-taut transitions. 
 \textbf{a}: Snapshots showing the human-payload disturbances.  
 \textbf{b}:  ES-CLF values.
 \textbf{c}: Payload velocity and position errors.  
 \textbf{d}: CBF values. \textbf{e}: Images of the payload observed by the onboard camera.
    \label{fig:manipulation_trajectory}}
    \vspace{-15pt}
\end{figure}

\subsubsection{Hybrid Mode Transitions During Trajectory Tracking} \label{subsec:traj_exp} 
In these experiments, we show that our method can satisfy the perception constraints and enforce the controller stability when the system is subjected to unexpected hybrid mode transitions during trajectory tracking tasks. 

Snapshots of the experiment are shown in Fig. \ref{fig:manipulation_trajectory}a where a disturbance pushes the payload upward, causing the cable to go slack, potentially moving the load outside the camera's FoV. The control performance, measured by the ES-CLF values and position errors, is shown in Fig. \ref{fig:manipulation_trajectory}b and \ref{fig:manipulation_trajectory}c. The results confirm that the controller ensures stability and convergence during the slack-to-taut transition while keeping the payload within the camera's FoV. The CBF values, $h(\vecx)$ in Fig. \ref{fig:manipulation_trajectory}d, show that despite disturbances, the perception constraint is maintained. We verify these results by observing that the camera images always contain the payload, as shown in Fig. \ref{fig:manipulation_trajectory}e. 

\subsubsection{Perception Safety During Dynamically Infeasible Trajectories} \label{subsubsec:exp_infeasible_traj}
In this experiment, we command the system to move to a position $1.5~\si{m}$ away in $1.5~\si{s}$, inducing a large acceleration, and causing the payload to exit the FoV. 
\begin{figure}[t]
\includegraphics[width=\columnwidth]{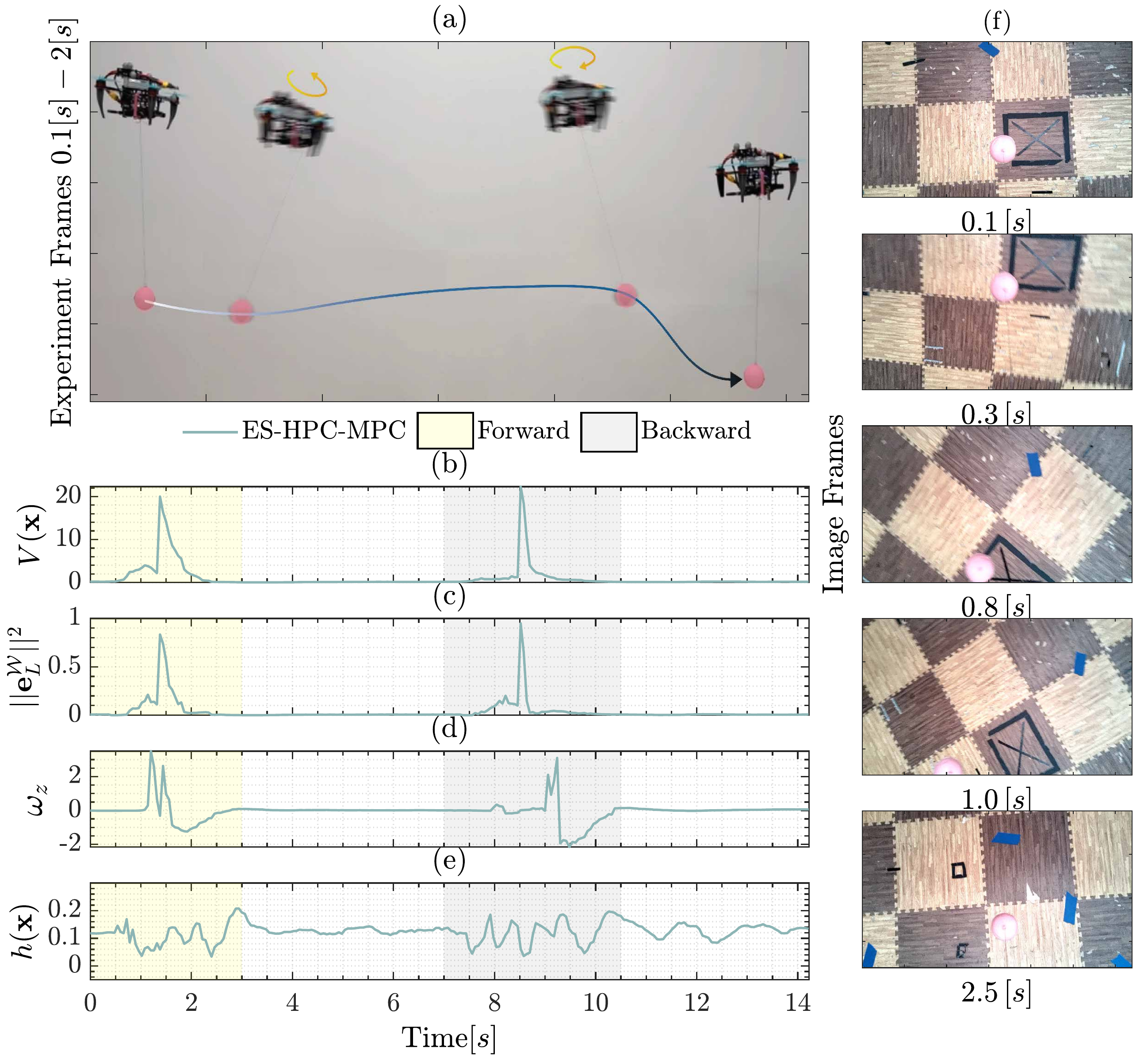}
    \caption{Tracking results for dynamically infeasible trajectory 
 \textbf{a}: Snapshots of the system movement.  
  \textbf{b}:  ES-CLF values.  
 \textbf{c}: Payload position errors.  
 \textbf{d}: Quadrotor angular velocity.  
 \textbf{e}: CBF values. \textbf{f}: Images of the payload observed by the camera.
    \label{fig:point_to_point}}
    \vspace{-15pt}
\end{figure}
As shown in Fig.
\ref{fig:point_to_point}a, the controller demonstrates a unique trade-off between trajectory tracking accuracy and payload visibility. Instead of reducing accelerations (and consequently the quadrotor's pitch angle) to maintain visibility, the controller leverages the camera's asymmetric resolution ($1280$ $\times$ $720$ pixels) to its advantage. The controller aligns the camera's wider dimension with the direction of motion by increasing   \(\omega_z\), as shown in Fig. \ref{fig:point_to_point}d. This motion improves the camera's visibility and ensures that the perception safety constraint is met Fig. \ref{fig:point_to_point}e. 

\begin{figure}[t]
\vspace{-5pt}
    \centering
    \includegraphics[width= \columnwidth]{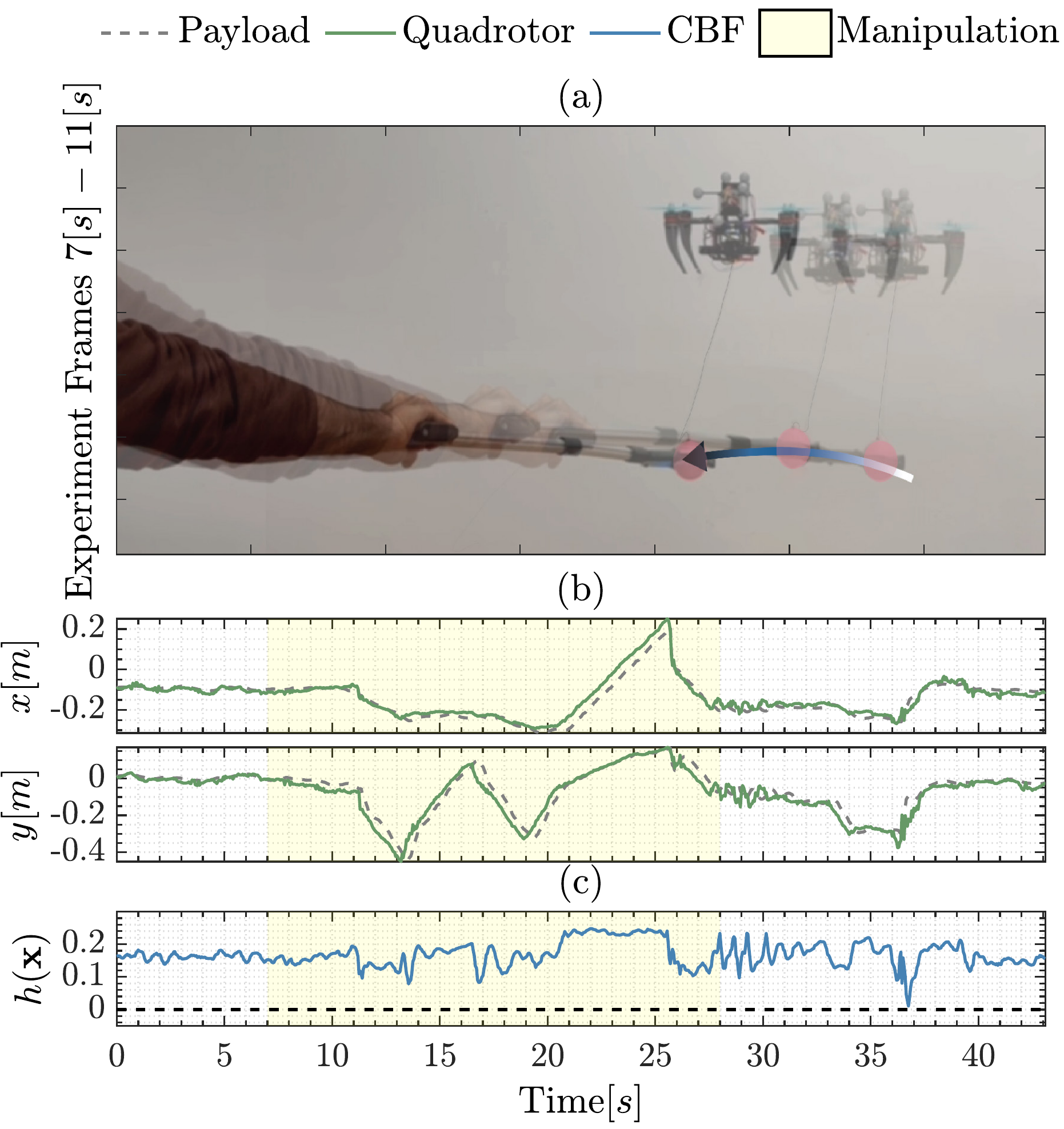} 
    \caption{Human-payload manipulation results. \textbf{a}: Snapshots showing movement of the system. \textbf{b}: Payload and quadrotor movements. \textbf{c}: CBF values throughout the experiment.\label{fig:manipulation_hover}}
        \vspace{-15pt}
\end{figure}

\subsubsection{Human-Payload Manipulation} \label{subsec:pampc_exp}
In this experiment, the system hovered around a fixed setpoint and we used a gripper to grasp the payload and move it around in all 3 directions. The quadrotor dynamically responds to the movements of the payload, exhibiting a faster reaction compared to the results presented in \cite{mrunal2025hpampc}. Notably, the video in \cite{mrunal2025hpampc} is shown at 3x playback speed, while our submitted video is presented at real-time 1x speed. We highlight this qualitative difference through a side-by-side comparison of both methods in our submitted video. The experimental results in Fig.~\ref{fig:manipulation_hover}, demonstrate the quadrotor's capability to maintain the payload in the FoV.
\vspace{-5pt}

\section{Conclusion} \label{sec:conclusion}
This work presents an Exponentially Stable Hybrid Perception-Constrained MPC for quadrotors with cable-suspended payloads, addressing challenges in stability, perception, and state estimation under hybrid dynamics using onboard sensors and compute. Our method enforces stability and maintains payload visibility within the camera’s FoV by leveraging the convergence guarantees of ES-CLFs and CBFs. Simulation and real-world experiments validate the effectiveness of our MPC method, demonstrating that our approach enforces stable trajectory tracking and perception awareness across various tasks without relying on external motion capture systems. In future work, we will investigate learning dynamic CLFs and CBFs that guarantee stability and safety perception for the quadrotor with cable-suspended payload directly from onboard images. We will also investigate sliding-mode techniques and online parameter estimation to enhance the controller’s adaptability under model mismatch.

\bibliographystyle{IEEEtran}
\bibliography{references.bib}

\end{document}